\documentclass[letterpaper, 10 pt, conference]{ieeeconf}  

\IEEEoverridecommandlockouts                              
\usepackage[latin1]{inputenc}
\usepackage{amsmath}
\usepackage{ntheorem}
\usepackage{amsfonts}
\usepackage{amssymb}
\usepackage{graphicx}
\usepackage{cleveref}
\usepackage{caption}
\usepackage{subcaption}
\usepackage{epstopdf}
\usepackage{algorithm}
\usepackage[noend]{algpseudocode}
\usepackage{soul,color}
\usepackage{tikz}
\usetikzlibrary{arrows}
\usetikzlibrary{positioning}

\newtheorem{definition}{Definition}
\newtheorem{proposition}{Proposition}
\newtheorem{corollary}{Corollary}
\newtheorem{observation}{Observation}
\newtheorem{theorem}{Theorem}

\theoremstyle{empty}

\begin{document}
	\title{\LARGE \bf Efficient Consensus-based Formation Control With Discrete-Time Broadcast Updates}
	\author{Fabio Molinari, J\"org Raisch
		\thanks{F. Molinari is with the Control Systems Group - Technische Universit\"at Berlin, Germany.}
		\thanks{J. Raisch is with the Control Systems Group - Technische Universit\"at Berlin, Germany \& Max-Planck-Institut f\"ur Dynamik Komplexer Technischer Systeme, Germany.}
		\thanks{\tt\small molinari@control.tu-berlin.de, 
			raisch@control.tu-berlin.de}
		\thanks{
			Authors want to thank Tim Berghoff Magnus for his contribution
			with the simulation environment.}
		\thanks{			
			This work was funded by the German Research Foundation (DFG) within their priority programme SPP 1914 "Cyber-Physical 
			Networking (CPN)". RA516/12-1.
		}
	}
	\maketitle
	\begin{abstract}
		This paper presents a consensus-based
		formation control strategy for
		autonomous agents moving in the plane
		with continuous-time single integrator dynamics.
		In order to save wireless resources (bandwidth, energy, etc),
		the designed controller exploits the
		superposition property of the wireless
		channel.
		A communication system, which is
		based on the Wireless Multiple Access Channel (WMAC) model
		and can deal with the presence of a fading channel
		is designed.
		Agents access the
		channel with simultaneous broadcasts
		at synchronous update times.
		A continuous-time
		controller 
		with discrete-time updates
		is proposed.
		A proof of convergence is given 
		and simulations are shown, 
		demonstrating
		the effectiveness of the suggested
		approach.
	\end{abstract}
\section{Introduction}
\label{sec:intro}
In the context of autonomous multi-agent systems,
coordination issues are of obvious importance.
Both the design of coordination
rules to achieve a desired overall
behaviour and the 
analysis of emergent behaviours for 
given local interaction rules have been widely
studied during the last two decades,
e.g. \cite{ren2007information,olfati2004consensus}.

One particular coordination problem 
for a multi-agent system
is \textit{formation control}.
A group of agents moving in space
are required to converge
towards a specific formation around 
a point, to be agreed on.
A distributed way 
for agents to achieve such a formation
is by executing a suitable consensus
protocol, e.g., see \cite{falconi2015edge}.
Consensus is a distributed
technique,
where agents have to agree 
on a variable of common interest by
exchanging information according
to a given communication topology.
Often, energy consumption
represents a critical aspect
in multi-agent scenarios (e.g., when agents are battery powered).
Then, the use of orthogonal channel access
methods for inter-agent communication,
which is standard,
is less than ideal.
In orthogonal channel access methods,
information exchange is agent-to-agent 
and avoids interference
by time or frequency multiplexing.
Instead,
the simultaneous broadcast
of information and the exploitation
of the wireless channel's superposition
property may be advantageous, see \cite{utschick2016communications,zheng2012fast, 
goldenbaum2012nomographic}.
A problem in the
context is that channel transmission properties
are unknown and time-varying.
This was addressed in 
\cite{molinari2018exploiting},
where a consensus protocol was suggested that can handle 
unknown channel coefficients.
The possible computational saving by using broadcast-based
protocol for so-called max-consensus
problems was quantified in \cite{molinari2018exploitingMax}.

In the remainder of this paper, 
a consensus-based formation control strategy is proposed.
Agents move in space with a continuous time dynamics
and can synchronously broadcast at discrete-time steps.
The proposed communication protocol makes use of the interference
of broadcast signals.
The formation control strategy takes inspiration from
the consensus protocol presented in 
\cite{almeida2012continuous}, where 
continuous-time dynamics and discrete-time communication
were brought together.


The paper is organized 
as follows. 
Section \ref{sec:notation} 
summarizes notation.
Section \ref{sec:probDesc}
defines the formation control problem.
A model of the wireless channel 
is presented in Section \ref{sec:WMAC},
and the communication system able to harness
the superposition property is designed in Section \ref{sec:commSys}.
Based on this, Section \ref{sec:propSol} presents
the consensus-based control strategy.
Convergence of the overall
scheme is given in Section \ref{sec:proof}.
A comparison with the standard
approach to the problem is in Section \ref{sec:iter}.
Simulation results are shown in Section \ref{sec:simul}.
Finally, Section \ref{sec:concl}
provides concluding remarks.

\section{Notation}
\label{sec:notation}
Throughout this paper,
$\mathbb{N}_0$, respectively $\mathbb{N}$,
denotes the set of nonnegative,
respectively positive, integers.
The sets of real numbers, nonnegative real numbers,
and positive real numbers are, respectively,
denoted $\mathbb{R}$, $\mathbb{R}_{\geq0}$, and $\mathbb{R}_{>0}$.
Given a set $\mathbb{S}$, its cardinality is
denoted by $|\mathbb{S}|$.

Given a matrix $A$ of dimension $n\times m$, 
the entry in position $(i,j)$
is $[A]_{ij}$. Matrix $A$ is positive (nonnegative)
if $[A]_{ij}>0$ ($[A]_{ij}\geq0$), $\forall i=1\dots n,\
\forall j=1\dots m$.
A nonnegative matrix $A$ of dimension $n\times n$ is called reducible if 
there exists a permutation matrix $Q$ such that
$
QAQ^{-1}
$
is of block upper-triangular form,
i.e. $A$ is reducible if 
and only if 
it can be brought into upper block-triangular 
form by identical row and column permutations.
If matrix $A$ is not reducible, is called irreducible.
A nonnegative square matrix $A$ of dimension $n$
is primitive if $\exists h\in\mathbb{N}$,
such that $A^h$ is positive.
Two nonnegative matrices $A$ and $B$ of the same dimension
are of the same type, and we write $A\sim B$,
if they have positive entries in the same positions.
$\mathbb{I}_n$ denotes the n-dimensional identity matrix,
$\mathbf{0}_{n\times m}$ is a matrix
of zeros with $n$ rows and $m$ columns, and
$\mathbf{1}_{n\times m}$ is a matrix
of ones with $n$ rows and $m$ columns.

A graph $\mathcal{G}$ is a pair $(\mathcal{N},\mathcal{A})$,
where $\mathcal{N}=\{1\dots n\}$ is the set of nodes
and $\mathcal{A}\subseteq\mathcal{N}\times\mathcal{N}$ set of arcs.
$(i,j)\in\mathcal{A}$ if and only if
an arc goes from node $i\in\mathcal{N}$
to node $j\in\mathcal{N}$.
Given a sequence of graphs 
constructed on the same set of nodes, i.e. 
$\{\mathcal{G}(k)\}_{k\in\mathbb{N}_0}$,
the set
$$\mathcal{N}_i(k):=\{j\in\mathcal{N} \mid (j,i)\in\mathcal{A}(k)  \} $$
contains the neighbors of agent $i$
in the graph $\mathcal{G}(k)$.
A path from node $i$ to node $j$
in graph $\mathcal{G}(k)$
is a sequence of arcs 
$$(l_0,l_1),(l_1,l_2),\dots,(l_{p-1},l_p)$$
with $p\geq1$,
$l_0=i$, $l_p=j$,
and $(l_i,l_{i+1})\in\mathcal{A}(k)$,
$\forall i=0\dots p-1$.
Graph $\mathcal{G}(k)$ is strongly connected if,
$\forall i,j\in\mathcal{N}$,
there exists a path from node $i$ to node $j$.
A weighted graph $\mathcal{G}(k)=(\mathcal{N},\mathcal{A}(k),\mathcal{W}(k))$,
with $\mathcal{W}(k)\in\mathbb{R}_{\geq0}^{n\times n}$,
is a graph in which every arc $(i,j)$
has a weight $[\mathcal{W}]_{ij}>0$.
Hence, $[\mathcal{W}]_{ij}=0$ implies that $(i,j)\not\in\mathcal{A}(k)$.
The weight matrix $\mathcal{W}(k)$
is also called adjacency matrix
of graph $\mathcal{G}(k)$.

Given $a,b\in\mathbb{R}$,
with $b>a$,
$\mathcal{U}(a,b)$ denotes
the uniform distribution between $a$ and $b$.

\section{Problem Description}
	\label{sec:probDesc}
	Let $\mathcal{N}=\{1,\dots,n\}$,
	$n\in\mathbb{N}$,
	be a set of autonomous agents
	moving in two-dimensional space
	with continuous-time dynamics and
	exchanging information
	over the wireless channel 
	at discrete-time
	steps. 
	

	The agents are assumed to exhibit
	single integrator dynamics, i.e.,
	$\forall i\in\mathcal{N}$,
	$\forall t\in\mathbb{R}_{\geq0}$,
	\begin{equation}
		\label{eq:dynamicsHolon}
		\dot{\mathbf{p}}_i(t)={\mathbf{u}}_i(t),
		\quad {\mathbf{p}}_i(0)={\mathbf{p}}_{i_0},
	\end{equation}	
	where
	${\mathbf{p}}_i(t)=[x_i(t),y_i(t)]'$ and
	$\mathbf{u}_i(t):=[u_i^x(t),u_i^y(t)]'$
	are the agents' state and control input,
	respectively.
	
	In a realistic framework,
	communication across the network 
	cannot be modeled as a continuous flow of information;
	instead, agents transmit and receive 
	data only at discrete update times
	$t_k\in\mathbb{R}_{\geq 0}$, $k\in\mathbb{N}_0$.
	In the following, 
	we assume that the interval
	$\Delta(k)=t_{k+1}-t_k$
	between two subsequent update times is
	bounded from below and above, i.e.
	$$\exists\Delta_1,\,\Delta_2\in\mathbb{R}_{\geq0}:\quad
	\Delta_1\leq\Delta(k)\leq\Delta_2,\,\forall k\in\mathbb{N}_0.$$
	The communication network 
	topology at update time $t_k$
	is modeled as a directed graph
	$\mathcal{G}(k)=(\mathcal{N},\mathcal{A}(k))$.
	
	The scope of this paper is
	to find a distributed control strategy
	such that the multi-agent system
	converges to a formation in the plane, i.e.,
	\begin{equation}
		\label{eq:formation}
		\forall i\in\mathcal{N},\
		\lim\limits_{t\rightarrow\infty}\mathbf{p}_i(t)=\mathbf{p}^* + \mathbf{d}_i,
	\end{equation}
	where $\mathbf{p}^*:=[{x}^*,{y}^*]'\in\mathbb{R}^2$ is the so-called
	\textit{centroid} of the formation and
	$\mathbf{d}_i:=[d^{x}_i,d^{y}_i]'\in\mathbb{R}^2$ is 
	the desired displacement of agent $i\in\mathcal{N}$
	from the centroid.
	As the aim of the distributed
	control system is to achieve
	consensus for $\mathbf{p}_i(t)-\mathbf{d}_i$,
	$\forall i\in\mathcal{N}$,
	it can be seen as a consensus
	protocol.

	As pointed out in Section \ref{sec:intro},
	the standard approach
	for implementing consensus protocols 
	is to use orthogonal channel access methods,
	which aim at providing each agent
	with all its neighbors' states.
	For the reasons outlined in Section \ref{sec:intro},
	we propose an alternative approach
	that is based on synchronous broadcasts
	of states,
	which allows for exploiting the wireless
	channel's superposition property.
\section{Communication System}
	The wireless channel is a shared broadcast medium;
	letting multiple users access 
	the same channel frequency spectrum simultaneously 
	results in interference, see \cite[pg. 100]{utschick2016communications}.
	Physically, the electromagnetic waves
	broadcast by a set of transmitters
	in the same frequency band
	superimpose at the receiver.
\subsection{Wireless Multiple Access Channel}
	\label{sec:WMAC}
	Consider the following scenario.
	A set of transmitting agents, 
	say $\underline{\mathcal{N}}=\{1,\dots,\underline{n} \}$,
	broadcast real-valued signals $\omega_i\in\mathbb{R}$,
	$i\in\underline{\mathcal{N}}$.
	Then, a simple {representation} known as 
	\textit{Wireless Multiple Access Channel (WMAC)},
	see e.g. \cite[Definition 5.2.1]{utschick2016communications},
	allows to model the value at a receiving agent.
	
	\begin{definition}[Wireless Multiple Access Channel (WMAC)]
		The WMAC
		between transmitters in $\underline{\mathcal{N}}$
		and a receiver is a map $\mathcal{W}:\mathbb{R}^{|\underline{\mathcal{N}}|}\mapsto\mathbb{R}$
		such that 
		\begin{equation}
			\label{eq:WMAC}
			\nu=
			\mathcal{W}(\{\omega_j\}_{j\in\underline{\mathcal{N}}})
			:=\sum_{j\in\underline{\mathcal{N}}} \xi_j\omega_j+\eta,
		\end{equation}
		where, $\forall j\in\underline{\mathcal{N}}$, $\xi_j\in\mathbb{R}$ is
		the (unknown) \textit{channel fading coefficient} between
		a transmitter $j$ and the destination,
		and $\eta$ is the receiver noise,
		see \cite{utschick2016communications}.
	\end{definition}
	In a collection of papers, 
	see \cite{goldenbaum2013harnessing,goldenbaum2009function,zheng2012fast} and also \cite{utschick2016communications},
	an \textit{ideal WMAC}, i.e.,
	$$\forall j\in\underline{\mathcal{N}},\ \xi_j=1\ \text{and}\ \eta=0,$$
	has been considered.
	\cite{molinari2018exploiting}
	considers a noiseless WMAC with \textit{power modulation},
	i.e.,
	\begin{equation}
		\label{eq:assumedMAC}
		\forall j\in\underline{\mathcal{N}},\ \xi_j\in\mathbb{R}_{>0} \text{ and } \eta=0,
	\end{equation}
	which will be assumed also throughout this paper.
%
	
\subsection{Communication System Design}
\label{sec:commSys}
	With the broadcast communication model (\ref{eq:WMAC})-(\ref{eq:assumedMAC})
	in hand, 
	it is possible to design a strategy that yields the desired
	results.
	At every update time $t_k$,
	$k\in\mathbb{N}_0$,
	each agent $j\in\mathcal{N}$
	broadcasts
	\begin{subequations}
		\label{eq:signals_sent}
		\begin{equation}
			\tau_j^x(k):=x_j(t_k)-d_j^x,
		\end{equation}
		\begin{equation}
			\tau_j^y(k):=y_j(t_k)-d_j^y.
		\end{equation}
	\end{subequations}
	Additionally, to handle
	the unknown channel coefficients
	in the WMAC model,
	the value
	\begin{equation}
		\tau_i'(k):=1
	\end{equation}
	is also broadcast.
	The three signals are
	broadcast orthogonally, i.e.,
	independent from each other
	(e.g., each signal is broadcast
	on a different frequency).
	By (\ref{eq:WMAC})-(\ref{eq:assumedMAC}),
	each agent $i\in\mathcal{N}$
	receives
	\begin{equation*}
		\nu_i^x(k)=\sum_{j\in \mathcal{N}_i(k)} \xi_{ij}(k) \tau_j^x(k),
	\end{equation*}
	\begin{equation*}
		\nu_i^y(k)=\sum_{j\in \mathcal{N}_i(k)} \xi_{ij}(k) \tau_j^y(k),
	\end{equation*}
	and
	\begin{equation*}
		\nu_i'(k)=\sum_{j\in \mathcal{N}_i(k)} \xi_{ij}(k) \tau_j'(k),
	\end{equation*}
	where
	$\xi_{ij}(k)\in\mathbb{R}_{>0}$
	is the (unknown) channel fading
	coefficient between transmitter
	$j\in\mathcal{N}_i(t_k)$
	and receiver $i$
	at update time $t_k$, $k\in\mathbb{N}_0$.	
	In the following, we will
	need the normalized values
	\begin{subequations}
		\begin{equation}
			\zeta_i^x(t_k):=\frac{\nu_i^x(k)}{\nu_i'(k)}
			=\frac{\sum_{j\in \mathcal{N}_i(k)} \xi_{ij}(k) (x_j(t_k)-d_j^x)}{\sum_{j\in \mathcal{N}_i(k)} \xi_{ij}(k)},
		\end{equation}		
		\begin{equation}
			\zeta_i^y(t_k):=\frac{\nu_i^y(k)}{\nu_i'(k)}
			=\frac{\sum_{j\in \mathcal{N}_i(k)} \xi_{ij}(k) (y_j(t_k)-d_j^y)}{\sum_{j\in \mathcal{N}_i(k)} \xi_{ij}(k)}.
		\end{equation}
	\end{subequations}
	In what follows, let,
	$\forall i,j\in\mathcal{N}$,
	$\forall k\in\mathbb{N}$,
	$h_{ij}(k)$ be 
	the normalized fading coefficient for
	broadcast transmission from $j$ to $i$
	at update step $k$,
	defined as
	\begin{align}
		\label{eq:normChCoeff}
		h_{ij}(k):=
		\begin{cases}
			\frac{\xi_{ij}(k)}{\sum_{j\in\mathcal{N}_i(k)}\xi_{ij}(k)}		&\text{if }(j,i)\in\mathcal{A}(k)\\
			0	&\text{else}
		\end{cases}		.
	\end{align}
	\begin{observation}
		\label{obs:chFadCoeffNorm}
		By construction, 
		$$\forall i,j\in\mathcal{N},\
		\forall k\in\mathbb{N}_0,\
		h_{ij}(k)\in[0,1].$$
		Moreover, 
		$\forall i\in\mathcal{N}$,
		$\forall k\in\mathbb{N}$,
		$$\sum_{j=1}^nh_{ij}(k)=\sum_{j\in\mathcal{N}_i(k)}h_{ij}(k)=1.$$
	\end{observation}
	
\section{Proposed solution}
	\label{sec:propSol}
	In the following, 
	since the dynamics
	in the $x$ and $y$ coordinates
	in (\ref{eq:dynamicsHolon})
	are decoupled,
	we analyze only the dynamics in $x$.
	An equivalent control strategy will be also applied
	to the $y$ dynamics.
	
	The following control
	strategy is inspired by 
	\cite{almeida2012continuous},
	where consensus is achieved in a network of 
	continuous-time agents with asynchronous 
	discrete-time updates. An extended consensus protocol
	is proposed here for achieving formation.
	Because we use broadcast transmission, 
	synchronism in transmission and update is required.
	
	We introduce 
	additional state variables
	$\vartheta^x_i(t)$, $i\in\mathcal{N}$, $t\in\mathbb{R}_{\geq0}$.
	While $x_i$ must be always continuous (it represents a position), 
	$\vartheta^x_i$ can have discontinuities at update times.
	The proposed control strategy is as follows:
	\begin{itemize}
		\item[$\blacksquare$] At update times $t_k,\ k\in\mathbb{N}_0$,
	\end{itemize}
	\begin{align}
		\label{eq:updateDyn}
		\begin{cases}
			x_i(t_k^+)=&x_i(t_k)\\
			\vartheta_i^x(t_k^+)=&(1-\sigma(t_k))\vartheta_i^x(t_k) + \sigma(t_k)d_i^x \\ 
			&+\sigma(t_k)\zeta_i^x(t_k)
		\end{cases}	
	\end{align}
	\begin{itemize}
		\item[$\blacksquare$] Between update times, i.e., $\forall t\in(t_k^+,t_{k+1}],\ k\in\mathbb{N}_0$,
	\end{itemize}
	\begin{align}
		\label{eq:dynamics}
		\begin{cases}
			\dot{x}_i(t)&=-a_i^x(t_k)(x_i(t)-\vartheta_i^x(t))\\
			\dot{\vartheta}_i^x(t)&=b_i^x(t_k)(x_i(t)-\vartheta_i^x(t))
		\end{cases}
	\end{align}
	where $t_k^+=\lim_{s\rightarrow0^+}t_k+s$ and 
	$\sigma(t_k)\in(0,1)$,
	$a_i^x(t_k),\ b_i^x(t_k)\in\mathbb{R}_{>0}$ are design parameters.
	Note that, in general,
	$\sigma(t_k)$,
	$a_i^x(t_k)$, and $b_i^x(t_k)$
	are going to be time-invariant.
	
	Between update times,
	the continuous time 
	dynamics (\ref{eq:dynamics})
	is executed,
	which reduces the absolute
	difference between the two state variables.
	In fact, if 
	$x_i(t)>\vartheta^x_i(t)$,
	then 
	$\dot{x}_i(t)<0$
	and $\dot{\vartheta}_i^x(t)>0$.
	Differently, if 
	$x_i(t)<\vartheta^x_i(t)$,
	we have 
	$\dot{x}_i(t)>0$
	and $\dot{\vartheta}_i^x(t)<0$.
	
	At every update time $t_k$, $k\in\mathbb{N}_0$,
	(\ref{eq:updateDyn}) is executed; it keeps the 
	positional variable $x_i$ constant,
	while it updates ${\vartheta}_i^x$. 
	The parameter $\sigma(t_k)$ is referred to as anti-stubbornness parameter; 
	in fact, the smaller $\sigma(t_k)$,
	the more agent $i$
	relies on its current value, 
	and the less on the received value
	$\zeta_i^x(t_k)$.
\section{Convergence}
	\label{sec:proof}	
	\begin{theorem}
		\label{prop:formationConv}
		Consider a set of communicating 
		agents with dynamics
		(\ref{eq:updateDyn})-(\ref{eq:dynamics}).
		If at every update time $t_k$, $k\in\mathbb{N}_0$,
		the network topology $\mathcal{G}(k)$ is strongly connected,
		then the system achieves the desired formation in the sense of (\ref{eq:formation}).
	\end{theorem}
	To prove Theorem \ref{prop:formationConv},
	first, define
	two additional state variables 
	$\tilde{x}_i(t)\in\mathbb{R}$ and $\tilde{\vartheta}^x_i(t)\in\mathbb{R}$
	as
	\begin{subequations}
		\label{eq:stateVarCetroid}
		\begin{equation}
			\tilde{x}_i(t):={x}_i(t)-d_i^x,
		\end{equation}
		\begin{equation}
			\tilde{\vartheta}^x_i(t):={\vartheta}^x_i(t)-d_i^x.
		\end{equation}
	\end{subequations}
	Clearly, 
	consensus in $\tilde{x}_i$ will imply (\ref{eq:formation}).
	Equations
	(\ref{eq:updateDyn})-(\ref{eq:dynamics})
	can be rewritten as
	\begin{itemize}
		\item[$\blacksquare$] At update times $t_k,\ k\in\mathbb{N}_0$,
	\end{itemize}
	\begin{equation}
		\label{eq:updateDyn_tilde}
		\begin{cases}
			\tilde{x}_i(t_k^+)&=\tilde{x}_i(t_k)\\
			\tilde{\vartheta}_i^x(t_k^+)&=(1-\sigma(t_k))\tilde{\vartheta}_i^x(t_k) + \sigma(t_k)\zeta_i^x(t_k)
		\end{cases}
	\end{equation}	
	\begin{itemize}
		\item[$\blacksquare$] Between update times, i.e., $\forall t\in(t_k^+,t_{k+1}],\ k\in\mathbb{N}_0$,
	\end{itemize}
	\begin{equation}
		\label{eq:dynamics_tilde}
		\begin{cases}
			\dot{\tilde{x}}_i(t)&=-a_i^x(t_k)({\tilde{x}}_i(t)-{\tilde{\vartheta}}_i^x(t))\\
			\dot{\tilde{\vartheta}}_i^x(t)&=b_i^x(t_k)({\tilde{x}}_i(t)-{\tilde{\vartheta}}_i^x(t))
		\end{cases}
	\end{equation}
	respectively, where
	\begin{equation}
		\zeta_i^x(t_k)={\sum_{j=1}^n h_{ij}(k)\tilde{x}_j(t_k)}.
	\end{equation}
	By (\ref{eq:dynamics_tilde}),
	$\forall i\in\mathcal{N}$,
	$\forall k\in\mathbb{N}_0$,
	\begin{equation}
		\label{eq:transition_update}
		\left[
			\begin{matrix}
				\dot{\tilde{x}}_i(t_{k+1})\\
				\dot{\tilde{\vartheta}}_i(t_{k+1})
			\end{matrix}
		\right]
			=
			\Phi_i(t_k)			
		\left[
			\begin{matrix}
				\dot{\tilde{x}}_i(t_{k}^+)\\
				\dot{\tilde{\vartheta}}_i(t_{k}^+)
			\end{matrix}
		\right]
	\end{equation}
	where the entries of the state transition matrix 
	\begin{align}
		\label{eq:stateTransMat}
		\Phi_i(t_k):&=
		\left[
			\begin{matrix}
				\Phi_i^a(t_k)	&\Phi_i^b(t_k)\\
				\Phi_i^c(t_k)	&\Phi_i^d(t_k)
			\end{matrix}
		\right],
	\end{align}
	are
	\begin{align}
		\Phi_i^a(t_k)&=\frac{a_i^x(t_k)e^{-(a_i^x(t_k)+b_i^x(t_k))(t_{k+1}-t_k^+)}+b_i^x(t_k)}
		{a_i^x(t_k)+b_i^x(t_k)},\\
		\Phi_i^b(t_k)&=\frac{a_i^x(t_k)(1-e^{-(a_i^x(t_k)+b_i^x(t_k))(t_{k+1}-t_k^+)})}
		{a_i^x(t_k)+b_i^x(t_k)},\\
		\Phi_i^c(t_k)&=\frac{b_i^x(t_k)(1-e^{-(a_i^x(t_k)+b_i^x(t_k))(t_{k+1}-t_k^+)})}
		{a_i^x(t_k)+b_i^x(t_k)},\\
		\Phi_i^d(t_k)&=\frac{b_i^x(t_k)e^{-(a_i^x(t_k)+b_i^x(t_k))(t_{k+1}-t_k^+)}+a_i^x(t_k)}
		{a_i^x(t_k)+b_i^x(t_k)}.
	\end{align}
	This can be intuitively seen 
	by observing that the matrix 
	\begin{align}
		A_i(t_k)=
		\left[
			\begin{matrix}
				-a_i^x(t_k) &a_i^x(t_k)\\
				b_i^x(t_k) &-b_i^x(t_k)				
			\end{matrix}
		\right]
	\end{align}
	has a \textit{zero} eigenvalue,
	while the second eigenvalue is
	$-(a_i^x(t_k)+b_i^x(t_k))$.
	Furthermore, 
	\begin{multline*}
		\Phi_i(t_k)=\\V_i(t_k)\text{diag}(1,\,e^{-(a_i^x(t_k)+b_i^x(t_k))(t_{k+1}-t_k^+)})V_i^{-1}(t_k),
	\end{multline*}
	where the columns
	of $V_i(t_k)$
	are right eigenvectors of $A_i(t_k)$
	corresponding to the two eigenvalues.
	\begin{observation}
		$\forall i\in\mathcal{N}$,
		$\forall k\in\mathbb{N}_0$,
		matrix $\Phi_i(t_k)$ is positive and row-stochastic
		by construction.
	\end{observation}
	This is easy to see as
	$a_i^x$, $b_i^x$,
	and $(t_{k+1}-t_k^+)$ are positive.
	Hence, $\Phi_i^a(t_k),\ \Phi_i^b(t_k),\ \Phi_i^c(t_k),\ \Phi_i^d(t_k) >0$.
	Furthermore,
	$\Phi_i^a(t_k)+\Phi_i^b(t_k)=\Phi_i^c(t_k)+\Phi_i^d(t_k)=1$.
	
	The system state vector
	(regarding movement in $x$-direction)
	is defined as
	$$\tilde{\mathbf{x}}(t):=
	\left[
		\tilde{x}_1(t),\dots,\tilde{x}_n(t),
		\tilde{\vartheta}_1(t),\dots,\tilde{\vartheta}_n(t)
	\right]^{'}
	 .$$
	The state evolution during each interval
	between $t_k^+$ and $t_{k+1}$
	can be described as
	\begin{equation}
		\label{eq:matForm_Phi}
		\tilde{\mathbf{x}}(t_{k+1})={\bf\Phi}(t_k)\tilde{\mathbf{x}}(t_{k}^+),
	\end{equation}
	where,
	$\forall k\in\mathbb{N}_0$,
	\begin{align}
		\label{eq:Phi}
		{\bf\Phi}(t_k)&:=
		\left[
			\begin{matrix}
				{\bf\Phi}^a(t_k) &{\bf\Phi}^b(t_k)\\
				{\bf\Phi}^c(t_k) &{\bf\Phi}^d(t_k)
			\end{matrix}
		\right]
		.
	\end{align}
	with
	\begin{align}
		{\bf\Phi}^a(t_k)=\text{diag}(\Phi_1^a(t_k)\dots\Phi_n^a(t_k)),\\
		{\bf\Phi}^b(t_k)=\text{diag}(\Phi_1^b(t_k)\dots\Phi_n^b(t_k)),\\
		{\bf\Phi}^c(t_k)=\text{diag}(\Phi_1^c(t_k)\dots\Phi_n^c(t_k)),\\
		{\bf\Phi}^d(t_k)=\text{diag}(\Phi_1^d(t_k)\dots\Phi_n^d(t_k)).
	\end{align}
	
	\begin{observation}
		$\forall k\in\mathbb{N}_0$,
		matrix ${\bf\Phi}(t_k)$ is nonnegative row-stochastic
		by construction.
	\end{observation}
	System (\ref{eq:updateDyn_tilde}) can be rewritten
	in matrix form as, $
	\forall k\in\mathbb{N}_0$,
	\begin{equation}
		\label{eq:matForm_D}
		\tilde{\mathbf{x}}(t_{k}^+)=D_n^\sigma(t_k)\tilde{\mathbf{x}}(t_{k}),
	\end{equation}
	where
	\begin{equation}
		\label{eq:Dn}
		D_n^\sigma(t_k)=
		\left[
			\begin{matrix}
				\mathbb{I}_n &\mathbf{0}_{n,n}\\
				\sigma(t_k)\mathfrak{A}(k)	&(1-\sigma(t_k))\mathbb{I}_n
			\end{matrix}
		\right].
	\end{equation}
	Matrix $\mathfrak{A}(k)$ 
	is the adjacency matrix associated with the
	directed graph $\mathcal{G}(k)$ with
	normalized fading coefficients as weights, i.e.,
	\begin{equation}
		\label{eq:adjacChannel}
		\forall i\in\mathcal{N},\ \forall j\in\mathcal{N},\
		[\mathfrak{A}(k)]_{ij}
		:=
		h_{ij}(k).
	\end{equation}
	\begin{proposition}
		\label{prop:adj}
		Given a strongly connected $\mathcal{G}(k)$, matrix
		$\mathfrak{A}(k)$ is nonnegative, irreducible, and row-stochastic.
		
		\begin{proof}
			By (\ref{eq:normChCoeff}), $\forall i,j\in\mathcal{N}$,
			$\forall k\in\mathbb{N}_0$, $h_{ij}(k)\in[0,1]$,
			then $\mathfrak{A}(k)$
			is nonnegative.
			By Observation \ref{obs:chFadCoeffNorm}, 
			$\sum_{j=1}^nh_{ij}(k)=1$, hence $\mathfrak{A}(k)$ is row-stochastic.
			Finally, by \cite[Thereom 6.2.24]{horn1990matrix},
			since $\mathcal{G}(k)$ is strongly connected, 
			$\mathfrak{A}(k)$ is also irreducible.
		\end{proof}
	\end{proposition}
	\begin{corollary}
		$\forall k\in\mathbb{N}_0$,
		$D_n^\sigma(t_k)$ is a nonnegative row-stochastic matrix.
		
		\begin{proof}
			Nonnegativity of $D_n^\sigma(t_k)$ follows
			directly from nonnegativity of $\mathfrak{A}(k)$
			and the fact that $\sigma(t_k)\in(0,1)$.
			It is also immediately clear that 
			each of the the first $n$ rows sums up to $1$.
			Let now be $l$ any value in $n+1\dots2n$. The $l$-th row sum of $D_n^\sigma(t_k)$ is
			\begin{align*}
				\sum_{j=1}^{2n}[D_n^\sigma(t_k)]_{lj}&=
				\sum_{j=1}^{n}[\sigma(t_k)\mathfrak{A}(k)]_{lj}
				+
				\sum_{j=1}^{n}[(1-\sigma(t_k))\mathbb{I}_n]_{lj}\\
				&=
				\sigma(t_k)\sum_{j=1}^{n}[\mathfrak{A}(k)]_{lj}
				+
				(1-\sigma(t_k))=1,
			\end{align*}
			as $\mathfrak{A}(k)=1$ is row-stochastic.
			This concludes the proof.			
		\end{proof}
	\end{corollary}
	\begin{proposition}
		\label{prop:fromShao}
		\cite[Prop. 1.2]{shao1985products}. If
		\begin{equation*}
			A=
			\left[
				\begin{matrix}
					A_1 &B\\
					C   &\mathbf{0}_{(n-l)\times(n-l)}
				\end{matrix}
			\right]
		\end{equation*}
		is an $n\times n$ nonnegative matrix, where $A_1$
		is an $l\times l$ ($1\leq l< n$) irreducible square matrix,
		and if $A$ contains no zero row or zero column, then $A$ is irreducible.
		
		\begin{proof}
			See \cite[proof of Prop. 1.2]{shao1985products}.
		\end{proof}
	\end{proposition}
	
	\begin{proposition}
		If $\mathcal{G}(k)$,
		$k\in\mathbb{N}_0$,
		is strongly connected,
		$\Omega(t_k):=\Phi(t_k)D_n^\sigma(t_k)$
		is irreducible and row-stochastic.
		
		\begin{proof}
			$\Omega(t_k)$ comes from the product of 
			two nonnegative row-stochastic matrices.
			By \cite{wolfowitz1963products}, $\Omega(t_k)$
			will also be nonnegative and row-stochastic.\par
			To prove its irreducibility, 
			note that (\ref{eq:Phi}) and (\ref{eq:Dn})
			imply
			\begin{equation*}
				\Omega(t_k)=
				\left[
					\begin{matrix}
						{\bf\Phi}^a(t_k)+\sigma(t_k){\bf\Phi}^b(t_k)\mathfrak{A}(k)		&(1-\sigma(t_k)){\bf\Phi}^b(t_k)\\
						{\bf\Phi}^c(t_k)+\sigma(t_k){\bf\Phi}^d(t_k)\mathfrak{A}(k)		&(1-\sigma(t_k)){\bf\Phi}^d(t_k)
					\end{matrix}
				\right].
			\end{equation*} 
			The product $C=AB$ of a diagonal matrix $A$ with 
			positive diagonal entries
			and a nonnegative irreducible
			matrix $B$ is an irreducible matrix $C$,
			since, by \cite[Pag. 30]{horn1990matrix},
			$C\sim B$
			and the irreducibility of a matrix depends only
			on its type (see \cite[pg. 735]{wolfowitz1963products}).
			By \cite[Theorem 1]{schwarz1966new}, 
			the sum of a nonnegative and an irreducible matrix
			is an irreducible matrix.
			From these two
			considerations, it immediately follows that 
			${\bf\Phi}^a(t_k)+\sigma(t_k){\bf\Phi}^b(t_k)\mathfrak{A}(k)$
			is irreducible. 
			
			Let's now consider matrix $\tilde{\Omega}(t_k)$, defined as
			\begin{equation*}
				\tilde{\Omega}(t_k):=
				\left[
					\begin{matrix}
						{\bf\Phi}^a(t_k)+\sigma(t_k){\bf\Phi}^b(t_k)\mathfrak{A}(k)		&(1-\sigma(t_k)){\bf\Phi}^b(t_k)\\
						{\bf\Phi}^c(t_k)+\sigma(t_k){\bf\Phi}^d(t_k)\mathfrak{A}(k)		&\mathbf{0}_{n\times n}
					\end{matrix}
				\right].
			\end{equation*} 
			By Proposition \ref{prop:fromShao}, 
			$\tilde{\Omega}(t_k)$ is irreducible. Let now
			$\bar{\Omega}(t_k)$ be a nonnegative matrix defined as
			\begin{equation*}
				\bar{\Omega}(t_k):=
				\left[
					\begin{matrix}
						\mathbf{0}_{n\times n}		&\mathbf{0}_{n\times n}\\
						\mathbf{0}_{n\times n}		&(1-\sigma(t_k)){\bf\Phi}^d(t_k)
					\end{matrix}
				\right].
			\end{equation*} 
			Hence, $\tilde{\Omega}(t_k)+\bar{\Omega}(t_k)={\Omega}(t_k) $. 
			Therefore,
			${\Omega}(t_k)$ is the sum
			of an irreducible and a nonnegative matrix;
			by \cite[Theorem 1]{schwarz1966new}, ${\Omega}(t_k)$ is irreducible.
		\end{proof}
	\end{proposition}
	\begin{corollary}
		If 
		$\Omega(t_k)$
		is irreducible, 
		it
		is also primitive.		
		
		\begin{proof}
			The irreducible matrix ${\Omega}(t_k)$ 
			has a positive diagonal by construction.
			By \cite[Theorem 1.4]{seneta2006non},
			any irreducible matrix with positive diagonal
			is also primitive.
			This concludes the proof.
		\end{proof}
	\end{corollary}
	By merging (\ref{eq:matForm_Phi}) and (\ref{eq:matForm_D}),
	the multiagent system 
	can be written as,
	\begin{equation}
		\label{eq:matform}
		\tilde{\mathbf{x}}(t_{k+1})=\Omega(t_k)\tilde{\mathbf{x}}(t_{k}),\quad
		k\in\mathbb{N}_0.
	\end{equation}
	\begin{proposition} 
		\label{prop:convTilde}
		If $\mathcal{G}(k)$ is strongly connected,
		$\forall k\in\mathbb{N}_0$,
		then the system (\ref{eq:matform})
		achieves consensus, i.e.,
		\begin{equation}
			\lim\limits_{t\rightarrow\infty}\tilde{\mathbf{x}}(t)=\tilde{\mathbf{x}}^*,
		\end{equation}
		where $\tilde{\mathbf{x}}^*={x}^*\mathbf{1}_{2n}$,
		${x}^*\in\mathbb{R}$.
		
		\begin{proof}
			Because of the assumptions
			on $\Delta(k)=t_{k+1}-t_k$,
			$$\lim\limits_{t\rightarrow\infty}\tilde{\mathbf{x}}(t)=
			\lim\limits_{k\rightarrow\infty}\tilde{\mathbf{x}}(t_k).$$
			By (\ref{eq:matform}),
			\begin{equation}
				\label{eq:whereConverges}
				\lim\limits_{k\rightarrow\infty}\tilde{\mathbf{x}}(t_k)=
				\lim\limits_{k\rightarrow\infty}{\bf\Omega}_{(0)}^{(t_{k-1})}\tilde{\mathbf{x}}(0),
			\end{equation}
			where ${\bf\Omega}_{(0)}^{(t_{k-1})}=
			\Omega(t_{k-1})\Omega(t_{k-2})\dots
			\Omega(t_{1})\Omega(t_{0})$. 
			By \cite{wolfowitz1963products,ren2007information}, 
			an infinite product of primitive row-stochastic 
			square matrices of dimension $2n$
			converges to
			\begin{equation}
				\label{eq:wolfowitz}
				\lim\limits_{k\rightarrow\infty}{\bf\Omega}_{(0)}^{(t_{k-1})}=
				\mathbf{1}_{2n}\tilde{\mathbf{v}}',
			\end{equation}
			where
			$\tilde{\mathbf{v}}\in\mathbb{R}_{>0}^{2n}$ and
			$\mathbf{1}_{2n}'\tilde{\mathbf{v}}=1$.
			By putting together (\ref{eq:whereConverges}) and (\ref{eq:wolfowitz}),
			\begin{equation}
				\label{eq:whereConverges_afterWolfowitz}
				\lim\limits_{t\rightarrow\infty}\tilde{\mathbf{x}}(t)=
				\lim\limits_{k\rightarrow\infty}\tilde{\mathbf{x}}(t_k)=
				\mathbf{1}_{2n}\tilde{\mathbf{v}}'\tilde{\mathbf{x}}(0)={x}^*\mathbf{1}_{2n},
			\end{equation}
			where ${x}^*=\tilde{\mathbf{v}}'\tilde{\mathbf{x}}(0)$.
		\end{proof}
	\end{proposition}

	By Proposition \ref{prop:convTilde},
	$\lim\limits_{t\rightarrow\infty}\tilde{\mathbf{x}}(t)={\mathbf{x}}^*$.
	This, in turn, by (\ref{eq:stateVarCetroid}),
	implies that, $\forall i\in\mathcal{N}$,
	\begin{equation}
		\label{eq:abscissa_conv}
		\lim\limits_{t\rightarrow\infty}{{x}}_i(t)={{x}}^*-d_i^x.
	\end{equation}
	As the dynamics
	in $x$- and $y$-directions
	are decoupled, as discussed in Section \ref{sec:propSol},
	the analogous result holds for the $y$-coordinate, i.e., 
	$\forall i\in\mathcal{N}$,
	\begin{equation}
		\label{eq:ordinate_conv}
		\lim\limits_{t\rightarrow\infty}{{y}}_i(t)={{y}}^*-d_i^y.
	\end{equation}
	This concludes the proof of Theorem \ref{prop:formationConv}.
\section{Number of transmissions}
	\label{sec:iter}
	As outlined in Section \ref{sec:commSys},
	the designed communication system requires
	$3$ orthogonal transmissions at every update time
	$t_k$, $k\in\mathbb{N}_0$.
	Clearly, the number of orthogonal transmissions does
	not depend on $\mathcal{G}(k)$.
	
	Consider now the case of implementing, instead,
	a standard orthogonal channel access method,
	e.g. TDMA (time-division multiple access) or FDMA
	(frequency-division multiple access).
	Since every node-to-node transmission
	needs to be orthogonal,
	at each update time $t_k$, $k\in\mathbb{N}_0$,
	the amount of required orthogonal 
	transmissions equals $g_k\in\mathbb{N}$,
	where
	\begin{equation}
		g_k:=2\sum_{i=1}^{n}|\mathcal{N}_i(t_k)|.
	\end{equation}
	In fact, each agent $i$ has to transmit, 
	at time $t_k$, $k\in\mathbb{N}_0$,
	to each neighbor $j\in\mathcal{N}_i(t_k)$,
	two orthogonal signals 
	(one for the $x$- and one for the $y$-dimension).
	
	In term of employed wireless resources,
	exploiting the broadcast property of the wireless
	channel
	results in a less expensive solution.
	In fact, 
	$g_k>3$ holds if
	$$\sum_{i=1}^{n}|\mathcal{N}_i(t_k)|>\frac{3}{2},$$
	which is always the case for 
	multi-agent systems with a strongly
	connected communication topology and $n>2$.

	An experimental comparison
	that takes into account also
	the effect of the fading channel is presented
	in the next section.
\section{Simulation}
	\label{sec:simul}		
	
	\begin{figure}[t]
		\includegraphics[width=\columnwidth]{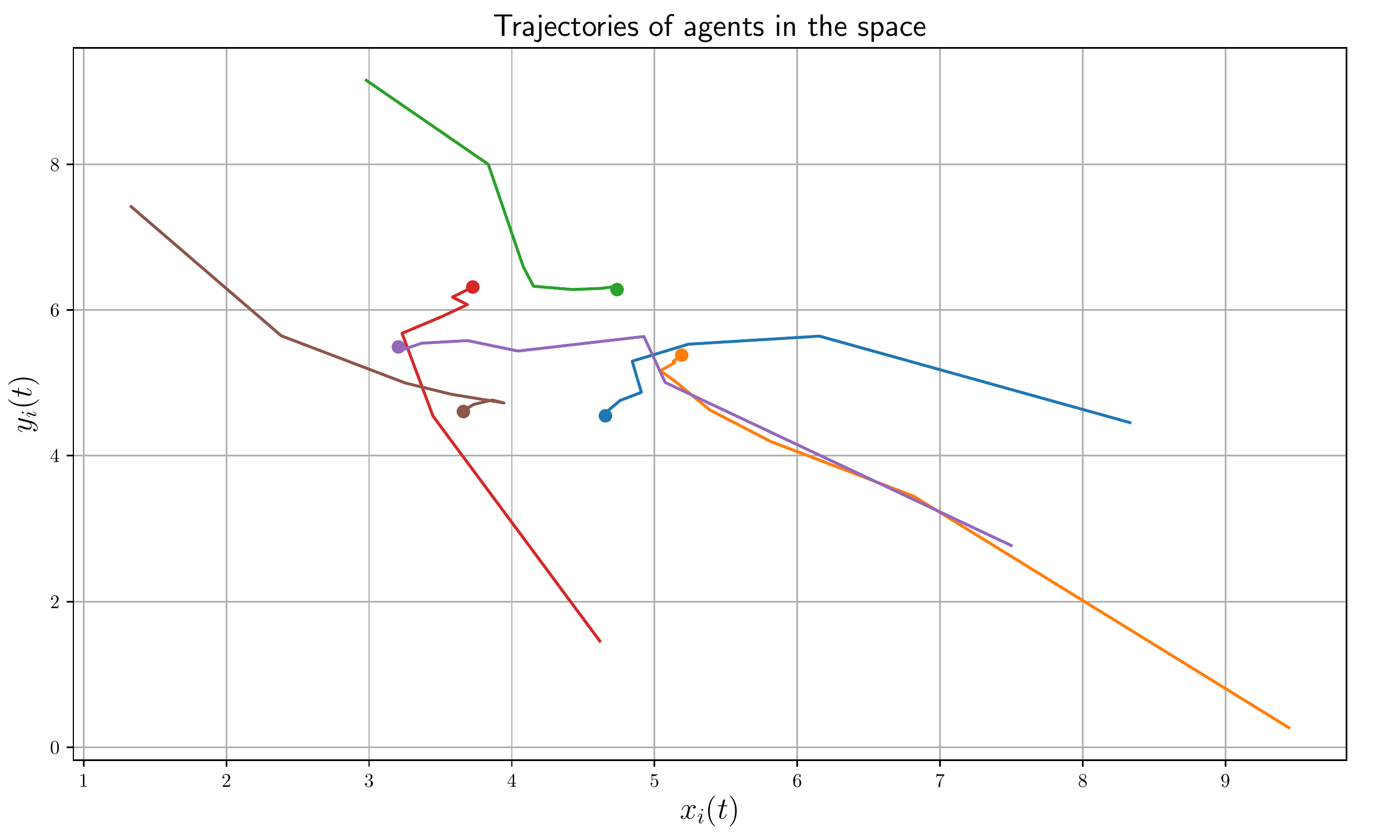}
		\caption{Trajectories in the space
		of agent in $\mathcal{N}$ seeking for a formation.
		The final achieved shape is a hexagon.}
				\label{fig:formation}
	\end{figure}

	\begin{figure}[t]
		\includegraphics[width=\columnwidth]{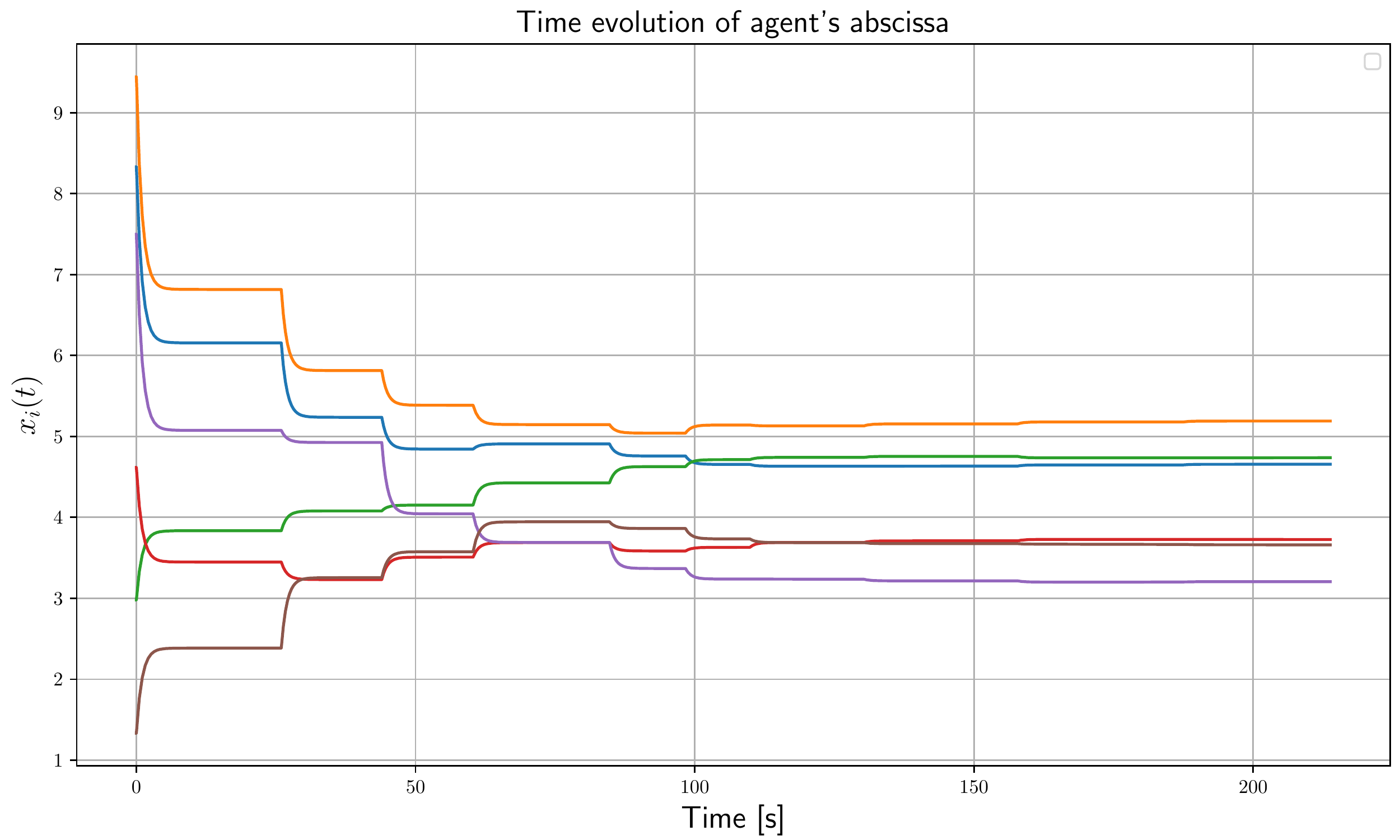}
		\caption{Evolution through time of the physical abscissa coordinate
		of each agent's dynamics.}
				\label{fig:abscissa}
	\end{figure}
	A set $\mathcal{N}$ composed of $n=6$ agents is given,
	where,
	$\forall i\in\mathcal{N}$, 
	$x_i(0)$ and $y_i(0)$
	are randomly chosen.
	The following parameters are used in the simulation,
	$\forall i\in\mathcal{N}$,
	$\forall k\in\mathbb{N}_0$,
	$a_i^x(t_k)=a_i^y(t_k)=0.5$,
	$b_i^x(t_k)=b_i^y(t_k)=0.5$,
	$\sigma_i(t_k)=0.8$.
	Desired displacements
	from the
	formation centroid ($d_i^x$ and $d_i^y$)
	are given, so that the final
	desired formation is a hexagon.
	
	A sequence of different strongly connected
	network topologies, i.e. $\{\mathcal{A}(k)\}_{k\in\mathbb{N}_0}$,
	is randomly chosen.
	At every update step $t_k$, $k\in\mathbb{R}_0$,
	also channel fading coefficients are randomly
	generated, so that,
	$\forall i\in\mathcal{N},\
	\forall j\in\mathcal{N}_i(k)$,
	the coefficients are
	independent and
	identically distributed,
	$\xi_{ij}(k)\sim\mathcal{U}(0,1)$.
	Also, the sequence $\{\Delta(k)\}_{k\in\mathbb{N}_0}$
	is randomly chosen, i.e.,
	$\forall k\in\mathbb{N},\
	\Delta(k)\sim\mathcal{U}(10,30)$.	
	
	Finally, simulation is run. 
	For solving the differential equations in (\ref{eq:dynamics}),
	the
	\textit{odeint} function from Python is used.
	In Figure \ref{fig:formation},
	the two-dimensional trajectories of agents
	are plotted. Clearly,
	they converge to the desired
	hexagonal formation.
	The evolution of the agents'
	$x$-position over time is shown in Figure \ref{fig:abscissa}.
	
	Also from Figure \ref{fig:abscissa},
	one can see that, at update time $t_7=130\, s$,
	agents have basically reached the desired formation.
	With regards to Section 
	\ref{sec:iter},
	reaching the goal
	has required $g_{bc}$ orthogonal transmissions,
	where 
	$g_{bc}=21$,
	i.e. $3$ orthogonal transmissions
	per $7$ update times.
	We compare this to the standard approach.
	Assume the worst case scenario
	which still guarantees that $\mathcal{G}(k)$
	is strongly connected, $\forall k\in\mathbb{N}_0$,
	i.e. $\sum_{i=1}^{n}|\mathcal{N}_i(t_k)|=10$, hence
	$g_k=20$. 
	This shows that, in the case at hand,
	the amount of 
	wireless resources used for converging
	with a broadcast-based protocol
	is equal to the amount of resources
	required for just one update step of
	the standard approach.
	The important amount of saved
	wireless resources
	explains why what presented in this paper
	constitutes an efficient formation control 
	strategy.


\section{Conclusion}
	\label{sec:concl}
	In this paper,
	a consensus-based formation control strategy
	has been proposed.
	It guarantees that agents with single integrator
	continuous-time dynamics achieve the
	desired formation,
	and it exploits the superposition property of the wireless channel.
	This translates in having synchronized discrete-time
	broadcasts.
	
	Note
	that the designed controller does not
	implement any collision avoidance strategy.
	This will be addressed in future work.
	Additionally, we aim
	to investigate
	how to extend (\ref{eq:dynamicsHolon})
	to a more general double-integrator dynamics.	
	Furthermore,
	we aim at experimentally validating the proposed protocol.

\bibliographystyle{IEEEtran}
\bibliography{paper_ECC19_Formation}

\end{document}